\documentclass[aps,twocolumn,prb,amsmath,amssymb]{revtex4}
\usepackage{graphicx}
\usepackage{epsfig}
\usepackage{color}
\usepackage{wasysym}

\usepackage{fancyhdr}

\begin{document}

\title{Finite-frequency noise in a non-interacting quantum dot}

\author{Redouane Zamoum}
\affiliation{Facult\'e des sciences et des sciences appliqu\'ees, Universit\'e de Bouira, rue Drissi Yahia, Bouira 10000, Algeria \
e-mail address: zamoum.redouane@gmail.com}

\author{Mireille Lavagna}
\affiliation{Univ. Grenoble Alpes, INAC-SPSMS, F-38000 Grenoble, France \
CEA, INAC-SPSMS, F-38000 Grenoble, France \
e-mail address: mireille.lavagna@cea.fr}

\author{Adeline Cr\'epieux}
\affiliation{Aix Marseille Universit\'e, Universit\'e de Toulon, CNRS, CPT UMR 7332, 13288 Marseille, France \
e-mail address: adeline.crepieux@cpt.univ-mrs.fr}

\begin{abstract}
We calculate the non-symmetrized finite-frequency NS-FF noise for a single-level quantum dot connected to reservoirs in the spinless non-interacting case. The calculations are performed within the framework of the Keldysh Green's function formalism in the wide band approximation limit. We establish the general formula for NS-FF noise for any values of temperature, frequency and bias voltage. The electron transfer processes from one to the other reservoir act via the  transmission amplitude and transmission coefficient depending on the energy. By taking the symmetrized version of this expression, we show that our result coincides with the expression of the finite frequency noise obtained by B\"uttiker using the scattering theory. We also give the explicit analytical expression for the NS-FF noise in the zero temperature limit. By performing numerical calculations, we finally discuss the evolution of the NS-FF noise spectrum when varying temperature, dot energy level, and coupling strength to the reservoirs, revealing a large variety of behaviors such as  different symmetry properties and changes of sign in the excess noise.

\end{abstract}

\maketitle

\section{Introduction}
Many recent experimental and theoretical studies are focused on the finite-frequency noise in nanoscopic systems. Conventionally, it is the symmetrized noise which is measured and calculated. This quantity is defined as the sum of two distinct spectral densities, wherein each of two contributions corresponds to a NS-FF noise, called emission noise and absorption noise respectively according to the sign of the frequency. In practice, it is more interesting to work with the NS-FF noise than with the symmetrized one. This is due to the fact that once the noise is symmetrized, it is no longer possible to separate the two non-symmetrized noise contributions. In other words, it is not possible to identify the absorption noise and the emission noise when one measures or calculates the symmetrized noise at finite frequency. Note that in many experiments, it is the excess noise which is measured. It corresponds 
to the difference between the noise at non-zero voltage and the noise at zero voltage, and its spectrum is symmetric in frequency in the absence of interaction. In Ref.~\onlinecite{billangeon06}, the two non-symmetrized contributions of the noise are measured separately in a Josephson junction. In the work by Lesovik 
and Loosen \cite{lesovik97}, a measurement scenario was proposed allowing to determine which spectral density is measured. In this scenario, the system is coupled to the detector in an inductive way: the noise is measured through the charge fluctuations inside the LC circuit. The NS-FF noise has been measured also for a quantum point contact \cite{zakka07} at frequencies in the range of 4-8~GHz. The emission and absorption noises have been measured separately using a superconductor-insulator-superconductor junction as a quantum detector, in the quantum regime of a superconducting resonant circuit at equilibrium \cite{basset10}. In another work, the emission noise was measured for a carbon nanotube quantum dot in the Kondo regime \cite{basset12}. 
\newline The study of the finite-frequency noise allows one to reach several properties of the system. In a carbon nanotube connected to leads, in which an electron is injected, the finite-frequency noise correlations allows one to calculate the charge transfer to the reservoirs \cite{lebedev05}. The NS-FF noises for a carbon nanotube and for a quantum wire have been calculated in Ref.~\onlinecite{safi08}. The frequency spectrum of the noise shows a strong asymmetry which is directly related to the ac-conductance \cite{safi09,joyez11,zamoum12}. 
\newline
\indent The calculation of the finite-frequency noise and its zero-frequency limit in a single-level quantum dot can be found in a certain number of works. In Ref.~\onlinecite{haug07}, the authors studied the symmetrized finite-frequency noise and its zero-frequency limit in such a system. The approach used was based on a non-equilibrium Green's function technique. In the same framework, the authors studied the different effects occurring in the resonant tunneling in a mesoscopic system connected to leads under the influence of time-dependent voltages\cite{jauho94}. In a more recent work, the zero frequency noise was calculated for a ferromagnet-quantum dot-ferromagnet system \cite{souza08} in the presence of Coulomb interaction and coherent spin-flip inside the dot. In a close context to the work developed here, the noise spectra for a quantum dot was studied in the case of finite bias voltage \cite{rothstein09}. Furthermore, in Ref.~\onlinecite{gabdank11}, the authors calculated the noise for a many-level quantum dot coupled to reservoirs and studied its dependence on the interaction strength, the coupling to the leads and the chemical potential. For a different system, the current-current correlators were derived for a tubular two-dimensional gas in the presence of Rashba spin-orbit interactions \cite{rothstein14}.
\newline 
\indent In this paper we study a single-level quantum dot connected to two reservoirs in the spinless non-interacting case. We focus on the calculation of the NS-FF noise in the left reservoir. In the Sec.~II, the model is presented in detail. The Hamiltonian and the different methods used in the calculation are explained. Then, in Sec.~III we expose the main result which is the expression of the NS-FF noise in the left reservoir and derive its limit cases. We confront the results that we have obtained with the formulas existing in the literature in some limit cases. In Sec.~IV, we plot the spectral density of the finite-frequency noise varying the parameters of the problem. We conclude in Sec.~V.


\section{Model}

\begin{figure}[h!]
\begin{center}
\includegraphics[width=7cm]{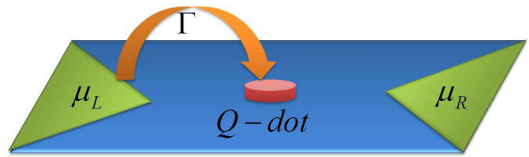}
\caption{Schematic representation of the single-level quantum dot connected to two reservoirs. The central region is supposed to be symmetrically coupled to the two reservoirs with a coupling strength $\Gamma$. $\mu_{L,R}$ are the chemical potential of the left and right reservoirs.}
\label{figure0}
\end{center} 
\end{figure}

We consider a single-level quantum dot as depicted in Fig.~\ref{figure0}. The Hamiltonian is the sum of four contributions, $H=H_L+H_R+H_T+H_{dot}$, which are respectively the Hamiltonian of the left and right reservoirs, the transfer Hamiltonian and the Hamiltonian of the dot. These expressions read as
\begin{eqnarray}
&&H_{\alpha=L,R}=\sum_{k;\alpha=L,R} \varepsilon_{k,\alpha} c_{k,\alpha}^{\dag}c_{k,\alpha}~,\\
&&H_T=\sum_{k;\alpha=L,R}V_{k,\alpha} c_{k,\alpha}^{\dag} d+h.c.~,\\
&&H_{dot}=\varepsilon_0 d^{\dag}d~.
\end{eqnarray}
The $c_{k,\alpha}^{\dag}$ and $c_{k,\alpha}$ are respectively the creation and annihilation operators in the reservoirs. The $d^{\dag}$ and $d$ are respectively the creation and annihilation operators in the dot, $\varepsilon_{k,\alpha}$ is the energy of an electron  with momentum $k$ in the reservoir $\alpha$, $V_{k,\alpha}$ is the transfer matrix element between the corresponding states, and $\varepsilon_0$ is the dot energy level. In this work, we neglect the Coulomb interactions between electrons in the dot. This is valuable when at least one of the characteristic energies of the problem, i.e. dot energy, temperature, voltage or coupling between the dot and the reservoirs, is large in comparison to the Coulomb interactions $U$, so that the latter can be neglected. Typically $U$ is estimated at about of a few meV in semiconductor quantum dot \cite{kobayashi10} and carbon nanotube quantum dot \cite{basset12}. In the non-interacting case, it is not necessary to consider the spin degree of freedom.

To calculate the noise in such a system, the most popular approach is the scattering theory developed by B\"uttiker (see Ref.~\onlinecite{blanter00} for a review). Here, we use an alternative approach based on the use of the nonequilibrium Green's function technique \cite{haug07}. It has the advantage to give results identical to the ones of the scattering theory in the non-interacting limit and it can be used in the future as a starting point to treat interacting systems.
\newline
\indent We focus on the calculation of the finite-frequency noise in the left reservoir. The finite-frequency noise in the right reservoir and the cross-noise can also be calculated in the same way\cite{zamoum15}. Note that when both finite-frequency and non equilibrium are considered, the finite-frequency noises in the left and right reservoirs respectively differ from each other. The non-symmetrized finite frequency left noise is defined from the current-current correlation function such as
\begin{eqnarray}
S_{LL}(\omega)=\int_{-\infty}^{\infty} S_{LL}(t,t') e^{-i\omega (t-t')}d(t-t')~,
\end{eqnarray}
where
\begin{eqnarray}
\label{Stt'}
S_{LL}(t,t')= \langle \delta \hat{I}_L(t) \delta \hat{I}_L(t') \rangle~,
\end{eqnarray}
with $\delta \hat{I}_L(t)=\hat{I}_L(t)-\langle I_L\rangle$, $\langle I_L\rangle$ being the average left current, and $\hat{I}_L$, the current operator in the left reservoir defined as $\hat{I}_L(t)=-ed\hat{N}_L(t)/dt$, with $\hat{N}_L(t)$, the electron number operator in the left reservoir.
Within this definition of $S_{LL}(\omega)$, the emission spectrum is observed at positive frequency, whereas the absorption spectrum is observed at negative frequency. Note that if one considers an alternative definition of finite-frequency noise, i.e., $S_{LL}(\omega)=\int_{-\infty}^{\infty} \langle \delta \hat{I}_L(t) \delta \hat{I}_L(t') \rangle e^{i\omega (t-t')}d(t-t')$, the noise spectrum is reversed: emission would be obtained at negative frequency and absorption at positive frequency.

We now give the derivation of $S_{LL}(\omega)$ in the situation of a single-level quantum dot connected to two reservoirs. We emphasize that the first part of the demonstration below until Eq.~(\ref{giniran00}) included, is completely general and holds even in the presence of interactions in the dot. The current operator in the left reservoir is given by\cite{jauho94}
\begin{equation}\label{current}
\hat{I}_L(t)=\frac{ei}{\hbar} \sum_k \Big( V_{k,L} c_{k,L}^{\dag} d-V_{k,L}^{\ast} d^{\dag} c_{k,L} \Big)~.
\end{equation}
Substituting the later expression for the current operator in Eq.~(\ref{Stt'}), the noise $S_{LL}(t,t')$ can be written as
\begin{eqnarray}
S_{LL}(t,t')&=&\Big( \frac{e}{\hbar} \Big)^2 \sum_{kk'}\Big[ V_{k,L}V_{k,L} G_1^{cd,>}(t,t')\nonumber \\
&&-V_{k,L}V_{k,L}^{\ast} G_2^{cd,>}(t,t')-V_{k,L}^{\ast}V_{k,L} G_3^{cd,>}(t,t') \nonumber \\ 
&&+V_{k,L}^{\ast}V_{k,L}^{\ast} G_4^{cd,>}(t,t')  \Big]-\langle I_L \rangle^2~,
\end{eqnarray}
where $G_i^{cd,>}(t,t')$, with $i\in[1,4]$, are the greater two-particle Green's functions mixing reservoir and dot operators\cite{haug07} defined as
\begin{eqnarray}
&&G_1^{cd,>}(t,t')=i^2 \langle c_{k,L}^{\dag}(t)d(t)c_{k',L}^{\dag}(t') d(t') \rangle~, \\
&&G_2^{cd,>}(t,t')=i^2 \langle c_{k,L}^{\dag}(t)d(t)d^\dag(t') c_{k',L}(t')  \rangle~, \\
&&G_3^{cd,>}(t,t')=i^2 \langle d^\dag(t) c_{k,L}(t)c_{k',L}^{\dag}(t') d(t')  \rangle~, \\
&&G_4^{cd,>}(t,t')=i^2 \langle d^\dag(t) c_{k,L}(t)d^\dag(t') c_{k',L}(t')  \rangle~.
\end{eqnarray}
In order to describe the system under out of equilibrium conditions, it is more convenient to introduce the contour-ordered counterpart of $G_i^{cd,>}(t,t')$, denoted as $G_i^{cd}(\tau,\tau')$, where the time variables $\tau$ and $\tau'$ belong to the Keldysh contour \cite{keldysh65}. $G_i^{cd,>}(t,t')$ can then be viewed as the greater components of the corresponding contour-ordered two-particle Green's functions $G_i^{cd}(\tau,\tau')$. In the same way, we introduce the contour-ordered counterpart of $S_{LL}(t,t')$, denoted as $S_{LL}(\tau,\tau')$. We now need to rewrite the Hamiltonian in the interaction representation. Using a $\mathcal{S}$-matrix expansion,  one can rewrite the current-current correlator in the interaction representation, where \cite{haug07}:
\begin{equation}
 \mathcal{S}=\sum_{j=0}^{\infty}\frac{(-i)^j}{j!}\int_C d\tau_1...\int_C d\tau_j \langle T_C \tilde{H}_T(\tau_1)...\tilde{H}_T(\tau_j) \rangle~,
\end{equation}
where $\tilde{H}_T$ is the transfer Hamiltonian written in the interaction representation and $T_C$ is the time ordered operator along the Keldysh contour $C$. After some technical transformations, we find  
\begin{widetext}
\begin{eqnarray}\label{giniran00}
 &&S_{LL}(\tau,\tau')=\Big( \frac{e}{\hbar} \Big)^2 \Bigg( \sum_{k} \vert V_{k,L} \vert^2 \Big[g_{k,L}(\tau',\tau)G(\tau,\tau')+g_{k,L}(\tau,\tau')G(\tau',\tau) \Big]  \nonumber \\
&&+\sum_{kk'} \vert V_{k,L} \vert^2 \vert V_{k',L} \vert^2 \int d\tau_1 \int d\tau_2 \Big[-g_{k,L}(\tau_1,\tau)g_{k',L}(\tau_2,\tau') G_1^{dd}(\tau,\tau',\tau_1,\tau_2)+g_{k,L}(\tau_2,\tau)g_{k',L}(\tau',\tau_1) G_2^{dd}(\tau,\tau',\tau_1,\tau_2)  \nonumber \\ 
&&-g_{k,L}(\tau,\tau_1)g_{k',L}(\tau_2,\tau') G_3^{dd}(\tau,\tau',\tau_1,\tau_2)-g_{k,L}(\tau,\tau_1)g_{k',L}(\tau',\tau_2) G_4^{dd}(\tau,\tau',\tau_1,\tau_2) \Big] \Bigg)-\langle I_L \rangle^2~,
\end{eqnarray}
\end{widetext}
where $g_{k,L}(\tau,\tau')$ is the bare contour-ordered one-particle Green's functions in the left reservoir, $G(\tau,\tau')$ the contour-ordered one-particle Green's functions in the dot, and $G_i^{dd,>}(\tau,\tau',\tau_1,\tau_2)$ are the contour-ordered two-particle Green's functions in the dot defined such as \cite{haug07}
\begin{eqnarray}
G_1^{dd}(\tau,\tau',\tau_1,\tau_2)=i^2 \langle T_C d(\tau)d(\tau')d^\dag(\tau_1)d^\dag(\tau_2) \rangle ~,\\
G_2^{dd}(\tau,\tau',\tau_1,\tau_2)=i^2 \langle T_C d(\tau)d^\dag(\tau')d^\dag(\tau_1)d(\tau_2) \rangle ~,\\
G_3^{dd}(\tau,\tau',\tau_1,\tau_2)=i^2 \langle T_C d^\dag(\tau)d(\tau')d(\tau_1)d^\dag(\tau_2) \rangle ~,\\
G_4^{dd}(\tau,\tau',\tau_1,\tau_2)=i^2 \langle T_C d^\dag(\tau)d^\dag(\tau')d(\tau_1)d(\tau_2) \rangle ~.
\end{eqnarray}

Let us emphasize that Eq.~(\ref{giniran00}) has been obtained without using any approximation. It forms the starting point for the calculation of the NS-FF noise for a large variety of systems; it still holds in the presence of Coulomb interactions in the dot as we have stressed above. It expresses the noise as a function of the full two-particle Green's functions in the dot, $G_i^{dd,>}(\tau,\tau',\tau_1,\tau_2)$, which need to be calculated in the presence of both coupling to reservoirs and interactions.
\newline
\indent In the rest of paper, we restrict the discussion to the case of a non-interacting quantum dot. In this case the two-particle Green's functions $G_i^{dd,>}(\tau,\tau',\tau_1,\tau_2)$ can be factorized into a product of two one-particle Green's functions; this corresponds to the Hartree-Fock approximation which is exact in the absence of interaction \cite{haug07}. Thus
\begin{equation}\label{factor}
G_i^{dd}(\tau,\tau',\tau_1,\tau_2)=G(\tau,\tau_2)G(\tau',\tau_1)-G(\tau,\tau_1)G(\tau',\tau_2)~.
\end{equation}
The noise contains two parts: $S(\tau,\tau')=S_{dis}(\tau,\tau')+S_{conn}(\tau,\tau')$, a disconnected part which is equal to the square of the average current according to $S_{dis}(\tau,\tau')=\langle I_L \rangle^2$ and so cancels this term, and a connected part which contains fifteen contributions. Finally the result is
\begin{widetext}
\begin{eqnarray}
&&S_{LL}(\tau,\tau')=\Big( \frac{e}{\hbar} \Big)^2 \Bigg( \sum_{k} \vert V_{k,L} \vert^2 \Big[g_{k,L}(\tau',\tau)G(\tau,\tau')+g_{k,L}(\tau,\tau')G(\tau',\tau) \Big]  \nonumber \\
&&+\sum_{kk'} \vert V_{k,L} \vert^2 \vert V_{k',L} \vert^2 \int d\tau_1 \int d\tau_2 \Big[-g_{k,L}(\tau_1,\tau)g_{k',L}(\tau_2,\tau') G(\tau,\tau_2)G(\tau',\tau_1)+g_{k,L}(\tau_2,\tau)g_{k',L}(\tau',\tau_1) G(\tau,\tau')G(\tau_1,\tau_2)  \nonumber \\ 
&&+g_{k,L}(\tau,\tau_1)g_{k',L}(\tau_2,\tau')  G(\tau',\tau)G(\tau_1,\tau_2)-g_{k,L}(\tau,\tau_1)g_{k',L}(\tau',\tau_2)  G(\tau_2,\tau)G(\tau_1,\tau') \Big] \Bigg)-\langle I_L \rangle^2~.
\end{eqnarray}
\end{widetext}
\indent One needs then to replace the contour integrals along the Keldysh contour by real time integrals along the two branches of the contour. This corresponds to an analytic continuation \cite{langreth76}. As usual we label the two branches of the contour by
the index $\eta$, where  $\eta=+$ for the upper branch, and $\eta=-$ for the lower branch. For a function $f(\tau_1)$ for instance, we can write
\begin{equation}
 \int_C d\tau_1 f(\tau_1)=\sum_{\eta_1} \eta_1 \int_{-\infty}^{\infty} dt_1 f(t_1^{\eta_1})~.
\end{equation}
In the expression of the finite-frequency noise above, we encounter terms with the following structure
\begin{equation}
 C(\tau,\tau')=\int_K d\tau_1 A(\tau,\tau_1)B(\tau_1,\tau')~.
\end{equation}
Taking the analytic continuation, one gets
\begin{equation}
 C^{\eta',\eta}(t',t)=\sum_{\eta_1} \eta_1 \int dt_1 A^{\eta',\eta_1}(t',t_1)B^{\eta_1,\eta}(t_1,t)~.
\end{equation}
The same procedure can be applied to the fifteen contributions to the noise. To illustrate the procedure, we write the result for one of these contributions
\begin{eqnarray}\label{T1}
 \mathcal{P}_1(t,t')&=&\Big( \frac{e}{\hbar} \Big)^2 \sum_{k} \vert V_{k,L} \vert^2 \Big[g_{k,L}^{<}(t',t)G^>(t,t') \nonumber \\
&&+g_{k,L}^{>}(t,t')G^<(t',t) \Big]~. 
\end{eqnarray}
In the same way, each of the fifteen contributions to the noise can be expressed in terms of the greater (or lesser) one-particle Green's functions, $g_{k,L}^{>,<}(t,t')$ and  $G^{>,<}(t,t')$.


\section{Results}
To get the expression of the NS-FF noise, more considerations are needed. First, a Fourier transform is performed on the fifteen contributions to the noise. As an illustration, we consider again here the Fourier transform of Eq.~(\ref{T1}) leading to
\begin{eqnarray}\label{tf1}
\mathcal{P}_1(\omega)&=& \frac{e^2}{h} \sum_{k} \vert V_{k,L} \vert^2 \int d\varepsilon \Big[ g_{k,L}^{<}(\varepsilon)G^>(\varepsilon-\omega) \nonumber \\
&&+g_{k,L}^{>}(\varepsilon-\omega)G^<(\varepsilon) \Big]~.
\end{eqnarray}
The next step is to incorporate the expressions of the bare one-particle Green's functions in the left reservoir\cite{mahan00}
\begin{eqnarray}
&&g_{k,L}^{>}(\varepsilon)=-2i\pi(1-n_L(\varepsilon_k))\delta(\varepsilon-\varepsilon_k)~,\\
&&g_{k,L}^{<}(\varepsilon)=2i\pi n_L(\varepsilon_k) \delta(\varepsilon-\varepsilon_k)~,
\end{eqnarray}
where $n_L(\varepsilon)$ is the Fermi-Dirac distribution function in the left reservoir. We get
\begin{eqnarray}\label{I}
\mathcal{P}_1(\omega)=\frac{ie^2}{h}\int d\varepsilon&&\Big[n_L(\varepsilon)\Gamma_L(\varepsilon)[G^r(\varepsilon-\omega)-G^a(\varepsilon-\omega)] \nonumber \\
&&+n_L(\varepsilon)\Gamma_L(\varepsilon)G^<(\varepsilon-\omega)\nonumber \\
&&-G^<(\varepsilon)[1-n_L(\varepsilon-\omega)]\Gamma_L(\varepsilon-\omega)\Big]~, \nonumber \\
\end{eqnarray}
where $G^r(\varepsilon)$ and $G^a(\varepsilon)$ are the retarded and advanced one-particle Green's functions of the dot respectively, and $\Gamma_{\alpha}(\varepsilon)=2\pi \Sigma_k{\delta(\varepsilon-\varepsilon_k) \vert V_{k,\alpha} \vert^2}$ is the coupling strength to the $\alpha$ reservoir. 
\newline 
\indent In the wide flat band limit when the electronic density of states in the $\alpha$ reservoir is assumed to be independent of energy with an infinite bandwidth, and $V_{k,\alpha}$ is independent of $k$, $\Gamma_{\alpha}(\varepsilon)$ is independent of energy: $\Gamma_{\alpha}(\varepsilon)=\Gamma_{\alpha}$. Considering the case of symmetric couplings to reservoirs $L$ and $R$: $\Gamma_L=\Gamma_R=\Gamma$, and making use of the following relations \cite{haug07}
\begin{eqnarray}
&&G^<(\varepsilon)=i\Gamma G^r(\varepsilon)[n_L(\varepsilon)+n_R(\varepsilon)]G^a(\varepsilon)~,\\
&& G^r(\varepsilon)- G^a(\varepsilon)=-2i\Gamma G^r(\varepsilon) G^a(\varepsilon)~,
\end{eqnarray}
one finally gets
\begin{eqnarray}\label{symm}
&&S_{LL}(\omega)=\frac{e^2}{h}\int d\varepsilon \Bigg[ \mathsf{T}(\varepsilon-\omega)\Big[ 1-\mathsf{T}(\varepsilon) \Big]f_{LR}(\varepsilon,\omega)  \nonumber \\
&&+\mathsf{T}(\varepsilon-\omega)\mathsf{T}(\varepsilon)f_{RR}(\varepsilon,\omega) +\mathsf{T}(\varepsilon)\Big[1-\mathsf{T}(\varepsilon-\omega)\Big]f_{RL}(\varepsilon,\omega)  \nonumber\\ 
&&+\Big[ \mathsf{T}(\varepsilon-\omega)\mathsf{T}(\varepsilon)
+\vert  \mathsf{t}(\varepsilon)- \mathsf{t}(\varepsilon-\omega)  \vert^2 \Big] f_{LL}(\varepsilon,\omega)\Bigg]~,
\end{eqnarray}
where $f_{\alpha\beta}(\varepsilon,\omega)=n_\alpha(\varepsilon)[1-n_\beta(\varepsilon-\omega)]$, and $ \mathsf{t}(\varepsilon)$, $ \mathsf{T}(\varepsilon)$ are the transmission amplitude and transmission coefficient respectively, given by
\begin{eqnarray}\label{amp}
 \mathsf{t}(\varepsilon)&=&\frac{i\Gamma}{ \varepsilon-\varepsilon_0+i\Gamma}~,\\\label{trans}
 \mathsf{T}(\varepsilon)&=&\frac{\Gamma^2}{( \varepsilon-\varepsilon_0)^2+\Gamma^2}~.
\end{eqnarray}
To derive Eq.~(\ref{symm}), we have used the relation: $2\mathsf{T}(\varepsilon)=\mathsf{t}(\varepsilon)+\mathsf{t}(\varepsilon)^{\ast}$, which is automatically fulfilled in the non-interacting case where $ \mathsf{t}(\varepsilon)$ and $ \mathsf{T}(\varepsilon)$ are given by Eqs.~(\ref{amp}-\ref{trans}). We point out that the later relation connecting $\mathsf{T}(\varepsilon)$ and $\mathsf{t}(\varepsilon)$ is completely general and is still valid in the presence of interactions; it corresponds to the optical theorem arising from the unitarity of the scattering matrix. We emphasize that Eq.~(\ref{symm}) is a novel and important result that has been derived using the nonequilibrium Green's function technique. 
\newline 
We now confront our result with the results existing in the literature in some limit cases.
First, in the perturbation limit (second order with the coupling $\Gamma$), only the term involving one-particle Green's function given by Eq.~(\ref{tf1}) enters into account. According to Ref.~\onlinecite{roussel15}, the noise can be expressed in terms of the dc-current. Such a relationship is called the perturbative fluctuation-dissipation relation. We get
\begin{equation}
S_{LL}(\omega)=e\Big[ N(\varepsilon+\omega)I_L(\varepsilon+\omega)+(1+N(\varepsilon-\omega))I_L(\varepsilon-\omega) \Big]~,
\end{equation}
with $N(\varepsilon)$ the Bose-Einstein distribution function and $I_L$ is the current given by \cite{haug07}:
\begin{eqnarray}
I_L(\varepsilon)&=&\frac{e}{h}\int \frac{d\varepsilon}{2\pi}\sum_{k} \vert V_{k,L} \vert^2 \Big[ g_{k,L}^{<}(\varepsilon)G^>(\varepsilon) \nonumber \\
&&-g_{k,L}^{>}(\varepsilon)G^<(\varepsilon) \Big]~.
\end{eqnarray}
\newline
Second, we check that the fluctuation-dissipation theorem is verified at equilibrium when $k_B T\gg \{eV,\varepsilon_0\}$. Using the following relation between the ac-conductance $G(\omega)$ and the noise asymmetry 
\begin{equation}\label{Gac}
G(\omega)=\frac{S_{LL}(-\omega)-S_{LL}(\omega)}{2\hbar \omega}~,
\end{equation}
and after some manipulations, we find that our results satisfy the fluctuation-dissipation theorem connecting the noise to the ac-conductance at equilibrium in agreement with Ref.~\onlinecite{joyez11}
\begin{equation}
S_{LL}(\omega)=2\hbar \omega N(\omega)G(\omega)~.
\end{equation}
\newline

Third, we calculate the symmetrized noise which can be obtained from Eq.~(\ref{symm}) by taking $S_{LL}^{sym}(\omega)=[ S_{LL}(\omega)+S_{LL}(-\omega) ]/2$. We get
\begin{eqnarray}\label{butt}
&&S_{LL}^{sym}(\omega)=\frac{e^2}{h} \int d\varepsilon \Bigg[ \mathsf{T}(\varepsilon) \Big[ 1-\mathsf{T}(\varepsilon-\omega) \Big] F_{RL}(\varepsilon,\omega) \nonumber \\
&&+\mathsf{T}(\varepsilon-\omega)\mathsf{T}(\varepsilon) F_{RR}(\varepsilon,\omega)+\mathsf{T}(\varepsilon-\omega) \Big[ 1-\mathsf{T}(\varepsilon) \Big] F_{LR}(\varepsilon,\omega) \nonumber \\
&&+ \Big[\mathsf{T}(\varepsilon-\omega) \mathsf{T}(\varepsilon)+\vert  \mathsf{t}(\varepsilon)- \mathsf{t}(\varepsilon-\omega)  \vert^2 \Big] F_{LL}(\varepsilon,\omega)\Bigg]~,
\end{eqnarray}
where $F_{\alpha,\beta}(\varepsilon,\omega)= n_{\alpha}(\varepsilon)[1-n_{\beta}(\varepsilon-\omega)]+n_{\beta}(\varepsilon-\omega)[1-n_{\alpha}(\varepsilon)]$. 
This expression coincides with the B\"uttiker formula for the symmetrized finite frequency noise obtained within the scattering matrix theory \cite{buttiker92,blanter00}. However, it does not coincide with the expression given in Ref.~\onlinecite{nazarov09} in which the term $\vert  \mathsf{t}(\varepsilon)- \mathsf{t}(\varepsilon-\omega)  \vert$ is missing. This discrepancy disappears when the transmission coefficient and amplitude are both energy independent. Moreover, in the zero frequency limit, we all recover the B\"uttiker formula for the zero frequency noise \cite{blanter00}:
\begin{eqnarray}
&&S_{LL}(0)=\frac{e^2}{h} \int d\varepsilon \Bigg[ [n_L(\varepsilon)-n_R(\varepsilon)]^2\mathsf{T}(\varepsilon)\Big[ 1-\mathsf{T}(\varepsilon) \Big]\nonumber \\
&&+\Big[n_{L}(\varepsilon)[1-n_{L}(\varepsilon)]+n_{R}(\varepsilon)[1-n_{R}(\varepsilon)]  \Big] \mathsf{T}(\varepsilon) \Bigg]~.
\end{eqnarray}
Fourth, in the zero temperature limit, the functions $f_{\alpha\beta}(\varepsilon,\omega)$ appearing in Eq.~(\ref{symm}) become a window function leading to the following result
\begin{eqnarray}\label{T0}
&&S_{LL}(\omega)=\frac{e^2}{h} \Bigg[ \Theta(-\omega)\int_{\mu_R+\omega}^{\mu_R}\mathsf{T}(\varepsilon-\omega)\mathsf{T}(\varepsilon)d\varepsilon \nonumber \\
&&+\Theta(-\omega)\int_{\mu_L+\omega}^{\mu_L} \Big[ \mathsf{T}(\varepsilon-\omega)\mathsf{T}(\varepsilon)+\vert  \mathsf{t}(\varepsilon)- \mathsf{t}(\varepsilon-\omega)  \vert^2 \Big] d\varepsilon \nonumber \\
&&+\Theta(-eV-\omega)\int_{\mu_L+\omega}^{\mu_R}\mathsf{T}(\varepsilon)\Big[1-\mathsf{T}(\varepsilon-\omega)\Big]d\varepsilon \nonumber\\ 
&&+\Theta(eV-\omega)\int_{\mu_R+\omega}^{\mu_L}\mathsf{T}(\varepsilon-\omega)\Big[ 1-\mathsf{T}(\varepsilon) \Big]d\varepsilon\Bigg]~,
\end{eqnarray}
where $\Theta(\omega)$ is the Heaviside step function. The result in Eq.~(\ref{T0}) is in agreement with the one obtained by Hammer and Belzig \cite{hammer11} in the same $T=0$ limit. Notice that since we choose to work with an alternative definition of the noise and a slightly different convention for the voltage to what is used in Ref.~\onlinecite{hammer11}, both results differ by a minus sign in front of the variable $\omega$. As far as the absorption noise is concerned, the first two terms in the r.h.s. of Eq.~(\ref{T0}) are non-zero at negative frequency, whereas the third term is non-zero for $\omega<-eV$ (assuming a positive bias voltage $V$). Thus all of these three terms contribute to the absorption noise. Moreover the only contribution to the emission noise comes from the fourth term in the r.h.s. of Eq.~(\ref{T0}), which is non-zero when the frequency belongs to the interval $[0,eV]$. At zero temperature, the emission noise vanishes when $\omega>eV$.
\newline
The integrals in Eq.~(\ref{T0}) can be performed analytically at temperature equal to zero in the non-interacting case where the transmission amplitude and the transmission coefficient are given by Eqs.~(\ref{amp}) and (\ref{trans}).  The result is 
\begin{eqnarray}
S_{LL}(\omega)= \Theta(-\omega)S_0(\omega)+\sum_{\sigma=\pm 1} \Theta(\sigma eV-\omega)S_\sigma(\omega)~,\nonumber\\
\end{eqnarray}
with
\begin{eqnarray}
&&S_0(\omega)=\frac{e^2}{h}\frac{\Gamma}{4\Gamma^2+\omega^2}\sum_{\sigma'=\pm 1}\nonumber\\
&&\times\left[\sigma'\Gamma^2\arctan\left(\frac{\varepsilon_0+\frac{V}{2}-\sigma'\omega}{\Gamma}\right)\right.\nonumber\\
&&+\sigma'(\Gamma^2+\omega^2)\arctan\left(\frac{\varepsilon_0-\frac{V}{2}-\sigma'\omega}{\Gamma}\right)\nonumber\\
&&-\frac{\Gamma^3}{\omega}\ln\left(\frac{\Gamma^2+\left(\varepsilon_0+\frac{V}{2}+\sigma'\omega\right)^2}{\Gamma^2+\left(\varepsilon_0+\frac{V}{2}\right)^2}\right)\nonumber\\
&&\left.-\frac{\Gamma(\Gamma^2+\omega^2)}{\omega}\ln\left(\frac{\Gamma^2+\left(\varepsilon_0-\frac{V}{2}+\sigma'\omega\right)^2}{\Gamma^2+\left(\varepsilon_0-\frac{V}{2}\right)^2}\right)\right]~,
\end{eqnarray}
and
\begin{eqnarray}
&&S_\sigma(\omega)=\frac{e^2}{h}\left[\sigma\Gamma\arctan\left(\frac{\varepsilon_0+\frac{V}{2}}{\Gamma}\right)\right.\nonumber\\
&&\left.-\sigma\Gamma\arctan\left(\frac{\varepsilon_0-\frac{V}{2}+\sigma\omega}{\Gamma}\right)\right]\nonumber\\
&&+\sum_{\sigma'=\pm 1}\left[\sigma'\frac{\Gamma^3}{4\Gamma^2+\omega^2}\arctan\left(\frac{\varepsilon_0-\sigma\sigma'\frac{V}{2}}{\Gamma}\right)\right.\nonumber\\
&&+\sigma'\frac{\Gamma^3}{4\Gamma^2+\omega^2}\arctan\left(\frac{\varepsilon_0-\sigma\sigma'\frac{V}{2}+\sigma'\omega}{\Gamma}\right)\nonumber\\
&&\left.+\frac{\Gamma^4}{(4\Gamma^2+\omega^2)\omega}\ln\left(\frac{\Gamma^2+\left(\varepsilon_0+\sigma'\frac{V}{2}-\sigma'\sigma\omega\right)^2}{\Gamma^2+\left(\varepsilon_0+\sigma'\frac{V}{2}\right)^2}\right)\right]~,\nonumber\\
\end{eqnarray}
where we have chosen a 
symmetric profile for the chemical potentials on either side of the junction, i.e. $\mu_{L,R}=\pm V/2$. At finite temperature, the integrals in Eq.~(\ref{symm}) can not be performed analytically and one has to perform a numerical integration as done in the next section.
\newline 


\section{Spectral density of the finite-frequency noise}

In order to follow the evolution of NS-FF noise as a function of the different parameters, we choose to plot the non-symmetrized excess noise $\Delta S_{LL}(\omega,V)=S_{LL}(\omega,V)-S_{LL}(\omega,0)$ as a function of frequency for different values of temperature and coupling strength. 
\begin{figure}[h!]
\begin{center}
\includegraphics[width=6.5cm]{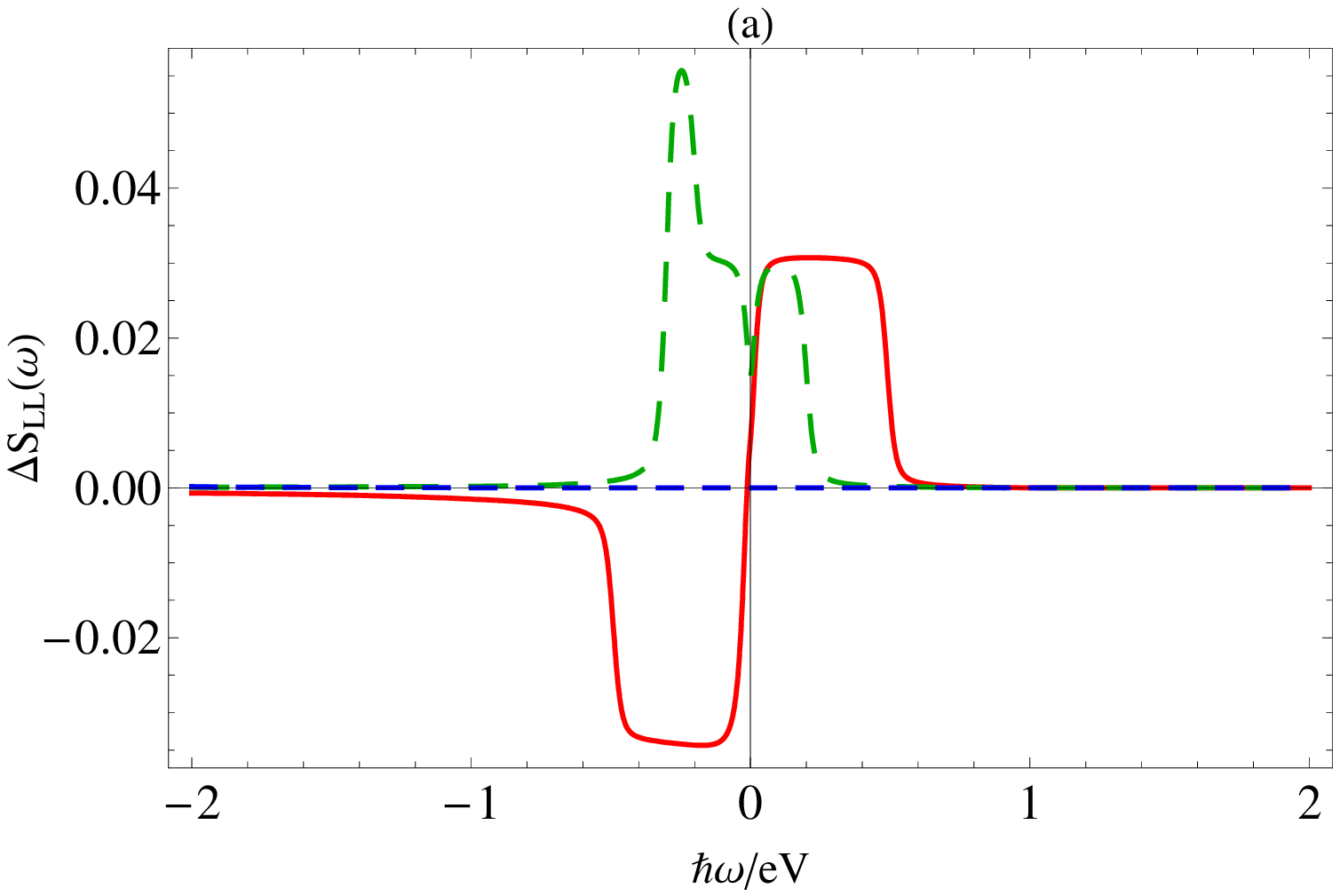}
\includegraphics[width=6.5cm]{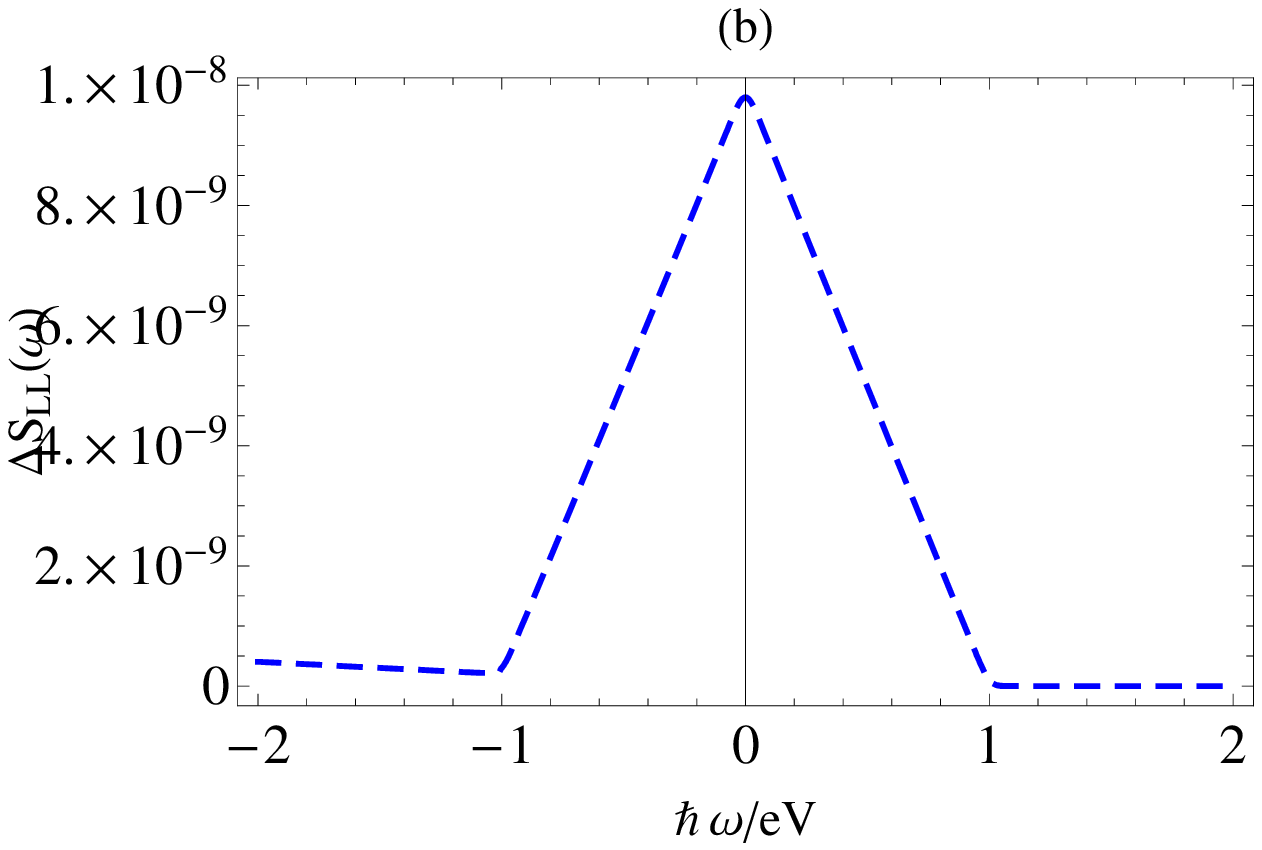}
\caption{(a) Spectral density of the non-symmetrized excess noise in units of $e^3V/\hbar$ at $k_BT/eV=0.01$ and $\Gamma/eV=0.01$. Solid red line corresponds to $\varepsilon_0/eV=0.01$, dashed green line to $\varepsilon_0/eV=0.3$ and dotted blue line to $\varepsilon_0/eV=100$; (b) Low-frequency zoom of the dotted blue line drawn in (a) for $\varepsilon_0/eV=100$.}
\label{figure1}
\end{center} 
\end{figure}
In Fig.~\ref{figure1}(a), we plot the spectral density of  the non-symmetrized excess noise at $k_BT/eV=0.01$ and $\Gamma/eV=0.01$ (low temperature and weak coupling regime) for different values of the dot energy level: $\varepsilon_0/eV=0.01$, $0.3$, and $100$. Fig.~\ref{figure1}(b) gives a low-frequency zoom of the solid red line drawn in Fig.~\ref{figure1}(a) for $\varepsilon_0/eV=100$. The effect of increasing $\varepsilon_0$ is to reduce the excess noise, resulting from the reduction of both $ \mathsf{t}(\varepsilon)$ and $ \mathsf{T}(\varepsilon)$  due to the presence of the $\varepsilon_0$ term in the denominators of the r.h.s. of Eqs.~(\ref{amp}) and (\ref{trans}). As can been seen in Fig.~\ref{figure1}(a), the spectrum profile of the excess noise drastically changes when increasing $\varepsilon_0$, going  from an anti-symmetric to an asymmetric and symmetric behaviors.  The excess noise presents a singularity in the vicinity of $\hbar\omega=\pm eV/2\hbar$ (for not too large 
values of $\varepsilon_0/eV$) and vanishes for frequency beyond $\hbar\omega=\pm eV/\hbar$ because the system can not emit at frequency larger than the voltage (in absolute value) and because the thermal noise contribution is negligible in the low temperature regime ($k_BT/eV=0.01$). For large values of $\varepsilon_0/eV$, the spectrum becomes fully symmetric in frequency, a result easily understood since the system is in the linear response regime when $eV$ is small.
The asymmetry observed in the excess noise spectrum for intermediate value of $\varepsilon_0/eV=0.3$ (dashed green line in Fig.~\ref{figure1}(a)) is characteristic of systems in a non-linear response regime\cite{safi09,joyez11}. The same profile has been found in the case of a one channel conductor coupled to a quantum of resistance \cite{zamoum12}. At low $\varepsilon_0/eV$, the spectrum is found to be anti-symmetric in frequency due to the choice of a symmetric profile of the chemical potentials on either side of the junction, i.e. $\mu_{L,R}=\pm V/2$. This dramatic change of the noise spectrum when varying the dot energy level was never noticed before. It is of importance for the ac-conductance since this quantity is related to the noise asymmetry through Eq.~(\ref{Gac}). In order to enlighten these latter results, we plot in Fig.~\ref{figure11} the non-symmetrized excess noise as a function of the dot energy level for different values of the coupling strength at small negative frequency. Here also we observe a change from negative sign to positive sign with increasing $\varepsilon_0$ providing that the coupling $\Gamma$ stays smaller than $eV$.

\begin{figure}[h!]
\begin{center}
\includegraphics[width=6.5cm]{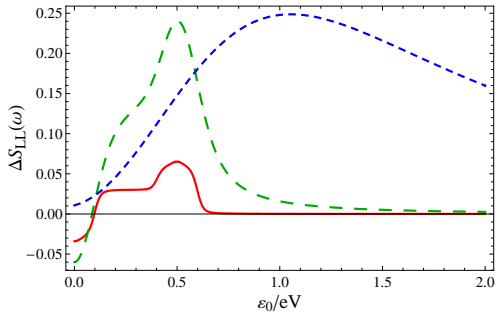}
\caption{Non-symmetrized excess noise in units of $e^3V/\hbar$ as a function of $\varepsilon_0/eV$ at $k_BT/eV=0.01$, and for fixed $\hbar \omega/eV=-0.1$ for different values of the coupling strength. Solid red line corresponds to $\Gamma/eV=0.01$, dashed green line to $\Gamma/eV=0.1$, and dotted blue line to $\Gamma/eV=1$. }
\label{figure11}
\end{center} 
\end{figure}

\begin{figure}[h!]
\begin{center}
\includegraphics[width=6.5cm]{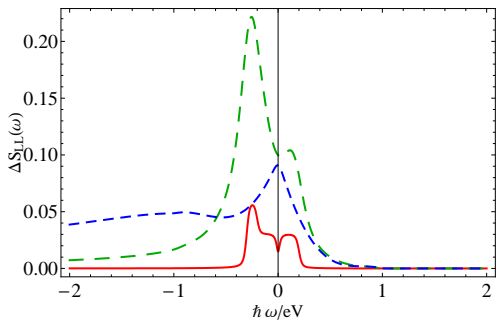}
\caption{Spectral density of the non-symmetrized excess noise in units of $e^3V/\hbar$ at $\varepsilon_0/eV=0.3$, and $k_BT/eV=0.01$. Solid red line corresponds to $\Gamma/eV=0.01$, dashed green line to $\Gamma/eV=0.1$ and dotted blue line to $\Gamma/eV=1$.}
\label{figure2}
\end{center} 
\end{figure}

\indent In Fig.~\ref{figure2}, we plot the spectral density of the non-symmetrized excess noise as a function of the frequency for different values of the coupling strength, at $\varepsilon_0/eV=0.01$ and $k_BT/eV=0.01$ (low temperature regime). In the weak coupling regime, the excess noise presents a singularity in the vicinity of $\hbar\omega=\pm eV/2$ and vanishes beyond those values. In the intermediate and strong coupling regimes, this singularity occurs instead at $\hbar\omega=\pm eV$ and the noise drops to zero at $\hbar\omega>eV$. Oppositely in the weak coupling regime the noise drops to zero at $\pm eV/2$.

\begin{figure}[h!]
\begin{center}
\includegraphics[width=6.5cm]{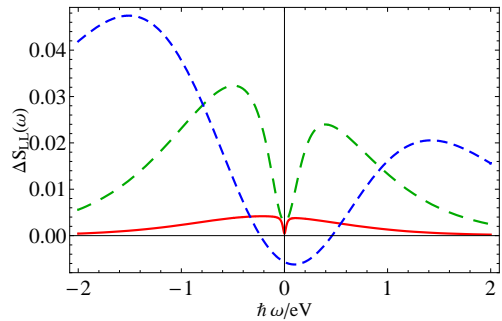}
\caption{Spectral density of the non-symmetrized excess noise in units of $e^3V/\hbar$ at $\varepsilon_0/eV=0.3$, and $k_BT/eV=0.5$. Solid red line corresponds to $\Gamma/eV=0.01$, dashed green line to $\Gamma/eV=0.1$ and dotted blue line to $\Gamma/eV=1$.}
\label{figure3}
\end{center} 
\end{figure}

\indent In Fig.~\ref{figure3}, we plot the spectral density of the non-symmetrized excess noise as a function of the frequency for different values of the coupling strength to the reservoirs at $k_BT/eV=0.5$ (high temperature regime). The singularities in the spectrum observed previously in the low temperature regime are no longer present in the high temperature regime. The obtained values for the excess noise are much higher than the values reported in Fig.~\ref{figure1} due to the thermal noise contribution. The noise spectrum is found to be symmetric in frequency in the low coupling regime (see solid red line); it becomes asymmetric in the intermediate and strong coupling regime (see dashed green and dotted blue lines). At high temperature and in the strong coupling regime, the excess noise may experience a sign change (see dotted blue line) meaning that the noise at zero voltage can exceed the noise at finite voltage due to the thermal noise contribution.


\section{Conclusion}
In this paper, we calculated and discussed the expression of NS-FF noise for a single-level quantum dot. We used the Keldysh formalism to evaluate the current-current correlator and then we performed a Fourier transform to get the expression of NS-FF noise. Our result is consistent with the B\"uttiker formula of the symmetrized finite-frequency noise obtained using the scattering theory since the symmetrization of our expression gives the formula obtained by B\"uttiker. Moreover, in the zero temperature limit, our result coincides with that of Hammer and Belzig. Varying temperature, dot level energy and coupling strengths, the profile of the excess noise spectrum reveals a rich behavior: it changes from a symmetric behavior, to an asymmetric or an anti-symmetric behavior. In the latter case, the excess noise exhibits a change of sign when the dot energy level varies.

The direct perspective of this work is the inclusion of Coulomb interactions which play an important role in low dimensional systems, in particular for the noise spectrum  \cite{blanter99} which is known to be more sensitive than the conductance. There exists several models to include Coulomb interactions, among them are the Luttinger liquid theory \cite{zamoum12,roussel15}, the interacting resonant level model \cite{boulat08a,boulat08b,branschadel10a,branschadel10b,freton14} and the real-time renormalization group method \cite{muller13,moca14}. The approach based on the nonequilibrium Green's function technique that we have presented here constitutes one of the possible ways to tackle this issue.

\acknowledgments
For financial support, the authors would like to thank the Indo-French Centre for the Promotion of Advanced Research (IFCPAR) under Research Project No.4704-02.



\begin{thebibliography}{99}
\bibitem{billangeon06}
P.-M.~Billangeon, F.~Pierre, H.~Bouchiat, and R.~Deblock, Phys. Rev. Lett {\bf 96}, 136804 (2006).

\bibitem{lesovik97}
G.B.~Lesovik and R.~Loosen. JETP Lett. {\bf 65}, 295 (1997).

\bibitem{zakka07}
E.~Zakka-Bajjani, J.~S\'egala, F.~Portier, P.~Roche, D.C.~Glattli, A.~Cavanna, and Y.~Jin, Phys. Rev. Lett {\bf 99}, 236803 (2007).

\bibitem{basset10}
J.~Basset, H.~Bouchiat, and R.~Deblock, Phys. Rev. Lett {\bf 105}, 166801 (2010).

\bibitem{basset12}
J.~Basset, A.Yu.~Kasumov, C.P.~Moca, G.~Zar\'and, P.~Simon, H.~Bouchiat, and R.~Deblock, Phys. Rev. Lett {\bf 108}, 046802 (2012).

\bibitem{lebedev05}
A.V.~Lebedev, A.~Cr\'epieux, and T.~Martin, Phys. Rev. B {\bf 71}, 047516 (2005).

\bibitem{safi08}
I.~Safi, C.~Bena, and A.~Cr\'epieux, Phys. Rev. B {\bf 78}, 205422 (2008).

\bibitem{safi09}
I.~Safi, arXiv:0908.4382 (2009); AIP Conf. Proc. {\bf 1129}, 431 (2009).

\bibitem{joyez11}
I.~Safi and P.~Joyez, Phys. Rev. B {\bf 84}, 205129 (2011).

\bibitem{zamoum12}
R.~Zamoum, A.~Cr\'epieux and I.~Safi, Phys. Rev. B {\bf 85}, 125421 (2012).

\bibitem{haug07} 
H.J.W.~Haug and A.P.~Jauho, in {\it Quantum kinetics in transport and optics of semiconductors}, Springer Series in Solid-State Sciences (2010).

\bibitem{jauho94} 
A.P.~Jauho, N.S.~Wingreen, and Y.~Meir, Phys. Rev. B {\bf 50}, 5528 (1994).

\bibitem{souza08}
F.M.~Souza, A.P.~Jauho, and J.C.~Egues, Phys. Rev. B {\bf 78}, 155303 (2008).

\bibitem{rothstein09}
E.A.~Rothstein, O.~Entin-Wohlman, and A.~Aharony, Phys. Rev. B {\bf 79}, 075307 (2009).

\bibitem{gabdank11}
N.~Gabdank, E.A.~Rothstein, O.~Entin-Wohlman, and A.~Aharony, Phys. Rev. B {\bf 84}, 235435 (2011).

\bibitem{rothstein14}
E.A.~Rothstein, B.~Horovitz, O.~Entin-Wohlman, and A.~Aharony, Phys. Rev. B {\bf 90}, 245425 (2014).

\bibitem{kobayashi10}T.~Kobayashi, S.~Tsuruta, S.~Sasaki, T.~Fujisawa, Y.~Tokura, and T.~Akazaki, Phys. Rev. Lett. {\bf 104}, 036804 (2010).

\bibitem{blanter00} 
Y.M.~Blanter and M.~B\"uttiker, Phys. Rep. {\bf 336}, 2 (2000).

\bibitem{zamoum15}
R.~Zamoum, M.~Lavagna, and A.~Cr\'epieux, arXiv:1511.08738 (2015).

\bibitem{keldysh65}
 L.V.~Keldysh, Sov. Phys. JETP {\bf 20}, 1018 (1965).
 
\bibitem{langreth76}
D.C. Langreth, in {\it Linear and nonlinear electron transport
in solids}, Vol. 17 of Nato Advanced Study Institute, Series~8: Physics, edited by J.T. Devreese and V.E. Van Doren
(Plenum, New York, 1976).

\bibitem{mahan00}
G.D.~Mahan, in {\it Many-particle physics}, Kluwer Academic/ Plenum publishers (2000).

\bibitem{roussel15}
B.~Roussel, P.~Degiovanni, and I.~Safi, arXiv:1505.02116 (2015).

\bibitem{buttiker92}
M.~B\"uttiker, Phys. Rev. B {\bf 45}, 3807 (1992).

\bibitem{nazarov09}
Y.V.~Nazarov and Y.M.~Blanter, in {\it Quantum Transport: Introduction to Nanoscience}, Cambridge University Press (2009).

\bibitem{hammer11}
J.~Hammer and W.~Belzig, Phys. Rev. B {\bf 84}, 085419 (2011).

\bibitem{blanter99}
Y.M.~Blanter and M.~B\"uttiker, Phys. Rev. B {\bf 59}, 10217 (1999).

\bibitem{boulat08a}
E.~Boulat and H.~Saleur, Phys. Rev. B {\bf 77}, 033409 (2008).

\bibitem{boulat08b}
E.~Boulat, H.~Saleur, and P.~Schmitteckert, Phys. Rev. Lett.
{\bf 101}, 140601 (2008).

\bibitem{branschadel10a}
A.~Bransch\"adel, E.~Boulat, H.~Saleur, and P.~Schmitteckert, Phys. Rev. Lett. {\bf 105}, 146805 (2010).

\bibitem{branschadel10b}
A.~Bransch\"adel, E.~Boulat, H.~Saleur, and P.~Schmitteckert, Phys. Rev. B {\bf 82}, 205414 (2010).

\bibitem{freton14}
L.~Freton and E.~Boulat, Phys. Rev. Lett. {\bf 112}, 216802 (2014).

\bibitem{muller13}
S.Y.~M\"uller, M.~Pletyukhov, D.~Schuricht, and S.~Andergassen, Phys. Rev. B {\bf 87}, 245115 (2013).

\bibitem{moca14}
C.P.~Moca, P.~Simon, C.-H.~Chung, and G.~Zar\'and, Phys. Rev. B {\bf 89}, 155138 (2014).

\end{thebibliography}
\end{document}